\documentclass[aps,preprint,prd,showpacs,nofootinbib]{revtex4}

\usepackage{latexsym}
\usepackage{amsmath}
\usepackage{graphicx}
\usepackage{subfigure}
\usepackage{dcolumn}
\usepackage{bm}
\usepackage{amssymb}
\usepackage{latexsym}
\usepackage{ulem}
\usepackage{color}
\usepackage[colorlinks,linkcolor=magenta,anchorcolor=cyan,citecolor=blue]{hyperref}

\def\be{\begin{equation}}
\def\ee{\end{equation}}
\def\ba{\begin{eqnarray}}
\def\ea{\end{eqnarray}}

\def\lf{\left}
\def\rt{\right}

\newcommand{\smallWidthLeft}{244pt}
\newcommand{\smallWidthRight}{224pt}

\begin{document}

\title{Towards the bounce inflationary gravitational wave }

\author{Hai-Guang Li$^{1}$\footnote{lihaiguang14@mails.ucas.ac.cn}}
\author{Yong Cai$^{1}$\footnote{caiyong13@mails.ucas.ac.cn}}
\author{Yun-Song Piao$^{1,2}$\footnote{yspiao@ucas.ac.cn}}

\affiliation{$^1$ School of Physics, University of Chinese Academy
of Sciences, Beijing 100049, China}

\affiliation{$^2$ Institute of Theoretical Physics, Chinese
Academy of Sciences, P.O. Box 2735, Beijing 100190, China}

\begin{abstract}

In bounce inflation scenario, the inflation is singularity-free,
while the advantages of inflation are reserved. We analytically
calculate the power spectrum of its primordial gravitational waves
(GWs), and show a universal result including the physics of the
bounce phase. The spectrum acquires a cutoff at large scale, while
the oscillation around the cutoff scale is quite drastic, which is
determined by the details of bounce. Our work highlights that the
primordial GWs at large scale may encode the physics of the bounce
ever happened at about $\sim 60$ efolds before inflation.

\end{abstract}

\maketitle

\section{Introduction}

The inflation
\cite{Guth:1980zm},\cite{Linde:1981mu},\cite{Albrecht:1982wi},\cite{Starobinsky:1980te}
is the paradigm of early universe. However, it is confronted with
the so-called ``initial singularity problem", since the inflation
itself is past-incomplete \cite{Borde:1993}. The solution to this
problem must involve something occurring before inflation. One
possibility is the so-called bounce inflation scenario
\cite{Piao:2003zm}, see also
\cite{Biswas:2013dry},\cite{Liu:2013kea},\cite{Falciano:2008gt},\cite{Mielczarek:2008pf},\cite{Xia:2014tda},\cite{Qiu:2014yn},\cite{Wan:2015hya},
in which initially the universe is contracting, and after the
bounce, the inflation starts.

Recently, the Planck collaboration \cite{Planck} have observed the
power deficit of CMB TT-mode spectrum at large scale. This
inspired theorists to think over the physics of the pre-inflation,
about $\sim 60$ efolds before the inflation  during which the
evolutions of largest scale perturbations are involved,
e.g.\cite{Cai:2015nya}. It is interesting that the
pre-inflationary physics suggested by the large-scale power
deficit might be relevant with the initial singularity problem.
The bounce universe, as the solution to this problem, has a long
history of study, see \cite{Battefeld},\cite{Lehners:2011} for
reviews. In
Refs.\cite{Piao:2003zm},\cite{Biswas:2013dry},\cite{Liu:2013kea},
it has been discovered that in the bounce inflation scenario the
large-scale anomalies of CMB TT spectrum may be explained
naturally. This achievement makes the bounce inflation acquire
increasing attention.

The primordial GWs is also the production of the inflation.
Recently, lots of experiments aiming at detecting GWs have been
implemented or will be implemented, which will bring us a new
epoch to understand the physics of inflation scenario,
e.g.\cite{Guzzetti:2016mkm}. There are also others designs to
explain the large-scale power deficit of scalar perturbation
spectrum, the suppression is attributed to the rapid rolling of
scalar field. However, the primordial GWs is independent of the
dynamics of scalar field, which thus may be used to identify the
physics of the pre-inflationary background.

Recently, the nonsingular bounce without modifying gravity has
been implemented, which may be ghost-free,
e.g.\cite{Qiu:2011cy},\cite{Easson:2011zy},\cite{Koehn:2013upa},
with possible embedding into supergravity
\cite{Khoury:2010gb},\cite{Koehn:2012te}. 
Thus the features of the primordial GWs at large scale might tell
us if such a bounce has ever happened at about $\sim 60$ efolds
before inflation. Moreover, it might also help us to speculate the
property of cyclic universe
\cite{Piao:2004me},\cite{Biswas:2008kj},\cite{Biswas},\cite{Banerjee}
with such a nonsingular bounce.

In this paper, we analytically calculate the bounce inflationary
GWs. The resulting spectrum is written with the recursive
Bogoliubov coefficients including the physics of the bounce phase.
We also show that our analytic result is completely consistent
with the numerical plotting for a realistic model of bounce
inflation.


\section{Overview of bounce inflation scenario}

\begin{figure}[htbp]
\centering
\includegraphics[scale=0.5]{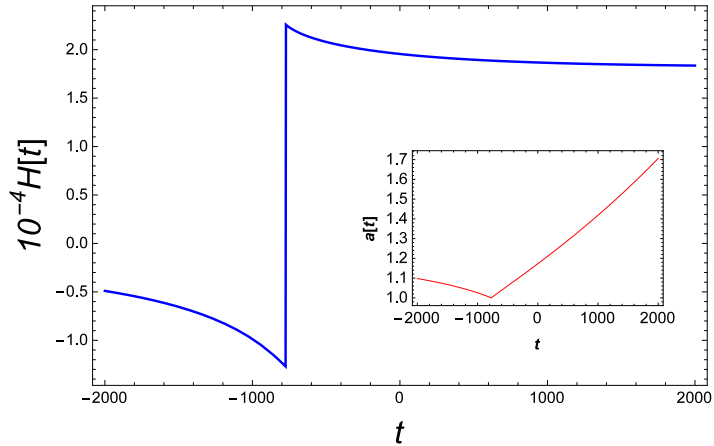}
\caption{ The evolution of $a$ and $H={\dot a}/a$ in bounce
inflation scenario, based on the model in Ref.\cite{Qiu:2014yn},
which will be showed in details in Sec.IV. }\label{scale1}
\end{figure}

In bounce inflation scenario, the inflation is singularity-free,
while the advantages of inflation are reserved, which simply and
naturally explains the universe we live.

The idea of bounce inflation showed itself first in
Ref.\cite{Piao:2003zm}, which explained the large-scale
suppression of scalar perturbation spectrum observed by WMAP.
Based on the Quintom bounce \cite{Cai:2007qw},
Ref.\cite{Cai:2007zv} investigated the evolution of the primordial
perturbations in details, and recently, Qiu and Wang
\cite{Qiu:2014yn}, and Wan et.al \cite{Wan:2015hya}, have
constructed a realistic model of bounce inflation without
instabilities by applying the higher-derivative operator.

The bounce inflation also has been implemented in the positively
curved background \cite{Falciano:2008gt}, or by modifying gravity
\cite{Biswas:2013dry}\cite{Mielczarek:2008pf}. The matching of
perturbation through the bounce with the modified gravity was
discussed in \cite{Biswas}.

Recently, Liu et.al \cite{Liu:2013kea} have found that the bounce
inflation brings not only the large-scale power deficit of CMB,
but also a large hemispherical power asymmetry, as implied by
Planck. Current bounds on the primordial GWs have been used to
constrain the bounce inflation \cite{Xia:2014tda}. However,
Ref.\cite{Xia:2014tda} used a step-like parameterisation of the
primordial GWs spectrum, based on \cite{Piao:2003zm}, which is
only a rough estimate missing the effect of the bounce. Recently,
the detecting of primordial GWs has been on the road, which will
possibly tell us more on the inflation and its origin. Therefore,
it is significant to have a full study for the bounce inflationary
GWs.

\section{Bounce inflationary GWs: analytical result }

Tensor perturbation $\gamma_{ij}$ satisfies $\gamma_{ii}=0$ and
$\partial_i \gamma_{ij}=0$. Its action is
 \be S_{\gamma}^{(2)}=\int d\eta d^3x {a^2\over
8}\lf[\lf(\frac{d {\gamma}_{ij}
}{d\eta}\rt)^2-(\vec{\nabla}{\gamma}_{ij})^2\rt]\,,
\label{action}\ee where $\eta=\int dt/a$ and $M_P^2=1$.

The Fourier series of $\gamma_{ij}$ is \be
h_{ij}(\eta,\mathbf{x})=\int \frac{d^3k}{(2\pi)^{3}
}e^{-i\mathbf{k}\cdot \mathbf{x}} \sum_{\lambda=+,\times}
\hat{h}_{\lambda}(\eta,\mathbf{k})
\epsilon^{(\lambda)}_{ij}(\mathbf{k}), \ee in which $
\hat{h}_{\lambda}(\eta,\mathbf{k})=
h_{\lambda}(\eta,k)a_{\lambda}(\mathbf{k})
+h_{\lambda}^*(\eta,-k)a_{\lambda}^{\dag}(-\mathbf{k})$,
polarization tensors $\epsilon_{ij}^{(\lambda)}(\mathbf{k})$
satisfy $k_{j}\epsilon_{ij}^{(\lambda)}(\mathbf{k})=0$,
$\epsilon_{ii}^{(\lambda)}(\mathbf{k})=0$,
 and $\epsilon_{ij}^{(\lambda)}(\mathbf{k})
\epsilon_{ij}^{*(\lambda^{\prime}) }(\mathbf{k})=\delta_{\lambda
\lambda^{\prime} }$, $\epsilon_{ij}^{*(\lambda)
}(\mathbf{k})=\epsilon_{ij}^{(\lambda) }(-\mathbf{k})$, the
annihilation and creation operators $a_{\lambda}(\mathbf{k})$ and
$a^{\dag}_{\lambda}(\mathbf{k}^{\prime})$ satisfy $[
a_{\lambda}(\mathbf{k}),a_{\lambda^{\prime}}^{\dag}(\mathbf{k}^{\prime})
]=\delta_{\lambda\lambda^{\prime}}\delta^{(3)}(\mathbf{k}-\mathbf{k}^{\prime})$.
The equation of motion for $u(\eta,k)$ is  \be
\frac{d^2u}{d\eta^2}+\left(k^2-\frac{a''}{a} \right)u=0,
\label{eom1} \ee where ${u}(\eta,k)={{ah_{\lambda}(\eta,k)}\over
{2}}$. The spectrum of primordial GWs is \be
P_T=\frac{k^3}{2\pi^2}\sum_{\lambda=+,\times} \lf|h_{\lambda}
\rt|^2=\frac{4k^3}{\pi^2}\cdot\frac{1}{ a^2} \lf|u \rt|^2, \quad
aH/k \gg 1.\label{pt} \ee

\begin{figure}[htbp]
\includegraphics[scale=0.25]{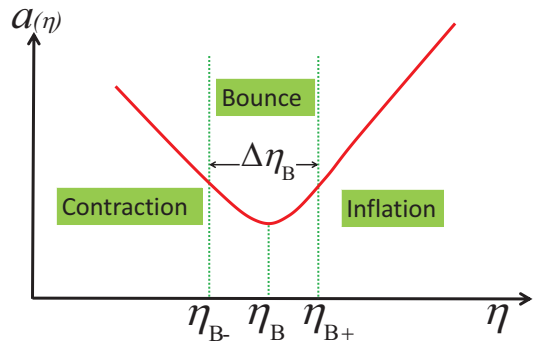}
\caption{ The illustration of the bounce inflation scenario. The
$\eta_{B-}$ is the beginning time of bounce phase, at which ${\cal
H}_{B-}=aH_{B-}<0$ and ${\dot H}_{B-}=0$. $\eta_{B}$ is the
so-called bounce point, at which $H=0$. $\eta_{B+}$ is the end
time of bounce phase, at which ${\cal H}_{B+}=aH_{B+}>0$ and
${\dot H}_{B+}=0$. After $\eta_{B+}$, the inflation
started.}\label{sketch}
\end{figure}


\subsection{The contracting phase }

The contracting phase is the evolution with $H<0$ and ${\dot
H}<0$. It ends at $\eta_{B-}$ when ${\dot H}=0$. Hereafter, ${\dot
H}>0$, the bounce starts.

The background can be parameterised as
 \be a_c(\eta)=a_{B-}\lf(\frac{\eta-\tilde{\eta}_{B-}}{\eta_{B-}-\tilde{\eta}_{B-}}\rt)^{\frac{1}{\epsilon_c-1}},\ee
  where $\tilde{\eta}_{B-}=\eta_{B-}-[(\epsilon_c-1){\cal
H}_{B-}]^{-1}$, noting the continuities of $a$ and $\cal H$ at
$\eta_{B-}$, and $\epsilon_c= -{{\dot H}/ H^2}$ and ${\cal
H}_{B-}$ is the comoving Hubble parameter at $\eta_{B-}$.
The initial state is Minkowski vacuum \be u_k=\frac{1}{\sqrt
{2k}}e^{i k \eta}.\ee Thus the solution of Eq.(\ref{eom1}) is \be
u_{k}={\sqrt{\pi\left|\eta-\tilde{\eta}_{B-}\right|}\over 2}
c_{1,1}H^{(1)}_{\nu_1}\left(k|\eta-\tilde{\eta}_{B-}|\right),\label{contract}\ee
where $\nu_{1}=\frac{\epsilon_c-3}{2(\epsilon_c-1)}$.

\subsection{The bounce phase}

The bounce phase is the evolution with ${\dot H}>0$. The Hubble
parameter is parameterised as \cite{Qiu:2014yn}\cite{Cai:2007zv}
\be H=\alpha(t-t_B),\label{bounceH}\ee with $\alpha M_P^2\ll 1$.
Thus we have \be a\simeq a_B e^{{1\over2} \alpha(t-t_B)^2}\simeq
a_B\left[1+{\alpha\over 2}(t-t_B)^2\right], \ee where $a=a_B$ at
$t=t_B$. Eq.(\ref{bounceH}) indicates that the bounce phase is
actually a superinflation phase with $H$ rapidly increasing. In
Ref.\cite{Biswas:2013dry}, it was argued that this phase results
in the large-scale power deficit in CMB. However, we will show
that in bounce inflation scenario this power deficit is actually
attributed to the contraction before the bounce.

The continuities of $a$ and $\cal H$ at $\eta_{B-}$ and
$\eta_{B+}$ suggest \be {\cal H}_{B+}={\cal H}_{B-} + \alpha a_B^2
\triangle\eta_B~. \label{HB}\ee We have ${\cal H}_{B+}={\alpha
a_B^2 \triangle\eta_B\over 2}$ for ${\cal H}_{B+}\simeq -{\cal
H}_{B-}$. ${\cal H}_{B+}$ is actually the comoving Hubble
parameter at the beginning time of the inflation. $\alpha$ and
$\triangle\eta_B=\eta_{B+}-\eta_{B-}$ encode the physics of the
bounce phase. Actually, since $a\simeq a_B$ during the bounce, we
can find $\eta-\eta_B=a_B^{-1}(t-t_B)$. Thus Eq.(\ref{eom1})
becomes \be u_k^{\prime\prime}+(k^2-\alpha a_B^2)u_k=0~. \ee Its
solution is \ba
\label{solbouncingsupten}u_k&=&c_{2,1}e^{l(\eta-\eta_B)}+c_{2,2}e^{-l(\eta-\eta_B)}~,
\ea where $l\equiv \sqrt{\alpha a_B^2-k^2}$.

\subsection{The inflationary phase}

The universe will inflate after $\eta_{B+}$. The background is
parameterised as
 \be a_{inf}(\eta)=a_{B+}\lf(\frac{\eta-\tilde{\eta}_{B+}}{\eta_{B+}-\tilde{\eta}_{B+}}\rt)^{\frac{1}{\epsilon_{inf}-1}}. \ee
Here, $\tilde{\eta}_{B+}=\eta_{B+}-[(\epsilon_{inf}-1){\cal
H}_{B+}]^{-1}$, noting the continuities of $a$ and $\cal H$ at
$\eta_{B+}$, and $\epsilon_{inf}= -{{\dot H}/ H^2}$. $H_{B+}={\cal
H}_{B+}/a$ sets the scale of inflation after the bounce,
$H=H_{B+}$.

Thus the solution of Eq.(\ref{eom1}) is
\ba u_{k}  = {\sqrt{\pi |\eta-\tilde{\eta}_{B+}|}\over 2}\left[
c_{3,1}H^{(1)}_{\nu_2}\left(k|\eta-\tilde{\eta}_{B+}|\right) +
c_{3,2}H^{(2)}_{\nu_2}\left(k|\eta-\tilde{\eta}_{B+}|\right)\right],\label{inflation}\ea
where $\nu_{2}=\frac{\epsilon_{inf}-3}{2(\epsilon_{inf}-1)}$.

\subsection{The spectrum of primordial GWs}

According to (\ref{pt}), if $\epsilon_{inf}=0$ we find \be
P_T=\frac{2H_{B+}^2}{\pi^2}|c_{31}-c_{32}|^2=P_{T}^{inf}|c_{31}-c_{32}|^2~,\label{pt2}
\ee where $P_{T}^{inf}=\frac{2H_{B+}^2}{\pi^2}$ is the standard
result of the slow-roll inflation.

The perturbation $\gamma_{ij}$ and its time derivative should be
continuous through the match surface. This suggests that we could
write the coefficients recursively as \ba \label{bb}\left(
\begin{array}{ccc} c_{3,1}\\c_{3,2}
\end{array}\right)&=&{\cal
M}^{(3,2)}\times{\cal
M}^{(2,1)}\times\left(\begin{array}{ccc}c_{1, 1}\\c_{1,
2}\end{array}\right), \label{rec1} \ea where the matric ${\cal
M}^{(2,1)}$ is \ba {\cal M}_{11}^{(2,1)}&=&e^{-l
y_1}\frac{\sqrt{\pi x_1}}{8l}\bigg\{k[-H^{(1)}_{\nu_{1}-1}(k
x_1)+H^{(1)}_{\nu_{1}+1}(k x_1)] +2[-\frac{ \nu_{1}}{x_1}+l+\alpha
a_B^2 y_1]H^{(1)}_{\nu_{1}}(k x_1)\bigg\},
\nonumber\\
{\cal M}_{12}^{(2,1)}&=&e^{-l y_1}\frac{\sqrt{\pi
x_1}}{8l}\bigg\{k[-H^{(2)}_{\nu_{1}-1}(k
x_1)+H^{(2)}_{\nu_{1}+1}(k x_1)] +2[-\frac{ \nu_{1}}{x_1}+l+\alpha
a_B^2 y_1]H^{(2)}_{\nu_{1}}(k x_1)\bigg\},
\nonumber\\
{\cal M}_{21}^{(2,1)}&=&e^{l y_1}\frac{\sqrt{\pi
x_1}}{8l}\bigg\{k[H^{(1)}_{\nu_{1}-1}(k x_1)-H^{(1)}_{\nu_{1}+1}(k
x_1)] +2[-\frac{ \nu_{1}}{x_1}+l-\alpha a_B^2
y_1]H^{(1)}_{\nu_{1}}(k x_1)\bigg\},
\nonumber\\
{\cal M}_{22}^{(2,1)}&=&e^{l y_1}\frac{\sqrt{\pi
x_1}}{8l}\bigg\{k[H^{(2)}_{\nu_{1}-1}(k x_1)-H^{(2)}_{\nu_{1}+1}(k
x_1)] +2[-\frac{ \nu_{1}}{x_1}+l-\alpha a_B^2
y_1]H^{(2)}_{\nu_{1}}(k x_1)\bigg\},\nonumber\label{matrix1} \ea
and ${\cal M}^{(3,2)}$ is \ba {\cal
M}_{11}^{(3,2)}&=&\frac{i\sqrt{\pi x_2}}{4}e^{l
y_2}\bigg\{k[H^{(2)}_{\nu_{2}-1}(k x_2)-H^{(2)}_{\nu_{2}+1}(k
x_2)] +2[\frac{ \nu_{2}}{x_2}+l-\alpha a_B^2
y_2]H^{(2)}_{\nu_{2}}(k x_2)\bigg\},
\nonumber\\
{\cal M}_{12}^{(3,2)}&=&\frac{i\sqrt{\pi x_2}}{4}e^{-l
y_2}\bigg\{k[H^{(2)}_{\nu_{2}-1}(k x_2)-H^{(2)}_{\nu_{2}+1}(k
x_2)] +2[\frac{ \nu_{2}}{x_2}-l-\alpha a_B^2
y_2]H^{(2)}_{\nu_{2}}(k x_2)\bigg\},
\nonumber\\
-{\cal M}_{21}^{(3,2)}&=&\frac{i\sqrt{\pi x_2}}{4}e^{l
y_2}\bigg\{k[H^{(1)}_{\nu_{2}-1}(k x_2)-H^{(1)}_{\nu_{2}+1}(k
x_2)] +2[\frac{ \nu_{2}}{x_2}+l-\alpha a_B^2
y_2]H^{(1)}_{\nu_{2}}(k x_2)\bigg\},
\nonumber\\
-{\cal M}_{22}^{(3,2)}&=&\frac{i\sqrt{\pi x_2}}{4}e^{-l
y_2}\bigg\{k[H^{(1)}_{\nu_{2}-1}(k x_2)-H^{(1)}_{\nu_{2}+1}(k
x_2)] +2[\frac{ \nu_{2}}{x_2}-l-\alpha a_B^2
y_2]H^{(1)}_{\nu_{2}}(k x_2)\bigg\}, \label{matrix2} \nonumber \ea
with the definitions $x_1=|\eta_{B-}-\tilde\eta_{B-}|$,
$y_1=(\eta_{B-}-\eta_B)$, $x_2=|\eta_{B+}-\tilde\eta_{B+}|$ and
$y_2=(\eta_{B+}-\eta_B)$.

\begin{figure}[htbp]
\centering
\includegraphics[scale=1]{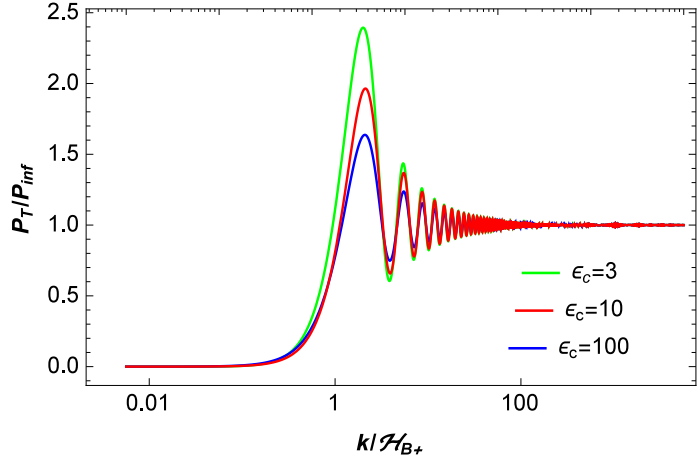}
\caption{ $P_T$ in Eq.(\ref{pt2}) for different $\epsilon_c$. We
set $\Delta\eta=0.2/{\cal H}_{B+}$ and
$\alpha=1.8\times10^{-3}{\cal H}_{B+}^2$. }\label{threep1}
\end{figure}

\begin{figure}[htbp]
\centering
\includegraphics[scale=0.7]{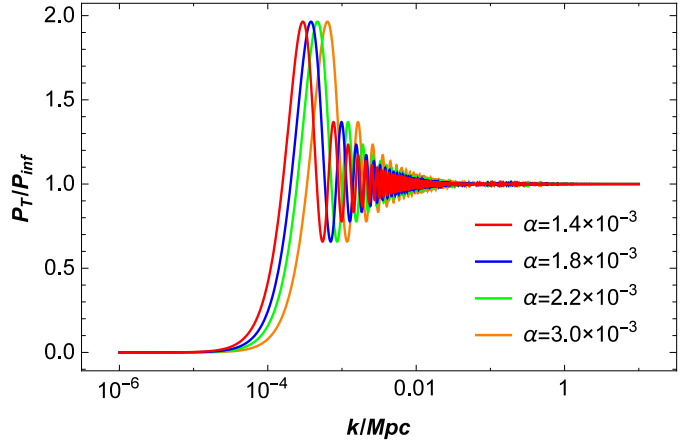}
\caption{ $P_T$ in Eq.(\ref{pt2}) for different $\alpha$ in unit
of ${\cal H}_{B+}^2$. We set $\epsilon_c=10$ and
$\Delta\eta=0.2/{\cal H}_{B+}$. }\label{threep2}
\end{figure}

The effects of pre-inflationary phases are encoded in ${\cal
M}^{(3,2)}$ and ${\cal M}^{(2,1)}$. Here, we set the slow-roll
parameter of inflation $\epsilon_{inf}\simeq 0$, thus $P_T$ is
only relevant with the parameters, $\alpha$, $\triangle\eta_B$,
and $\epsilon_c$, noting that $\epsilon_c\geq 3$ must be satisfied
to avoid the cosmic anisotropy problem \cite{Erickson:2003zm}. We
plot $P_T$ in Fig.\ref{threep1} and Fig.\ref{threep2} by altering
the values of different parameters.
We see that for $k>{\cal H}_{B+}$, $P_T\sim k^0$ with a damped
oscillation, and is blue shift for small $k$ modes. We may
analytically estimate it as follows.

For large $k$-modes, i.e. $k>{\cal H}_{B+}$, \ba
|c_{3,1}-c_{3,2}|^2 &\approx&1-(\frac{{\cal
H}_{B+}}{k}-\frac{\alpha a_B^2 \Delta\eta_B
}{2k})\sin(\frac{2k}{{\cal H}_{B+}})\nonumber\\
&+& (\frac{\epsilon_c{\cal H}_{B+}}{2k}-\frac{3{\cal
H}_{B+}}{2k}+\frac{\alpha a_B^2 \Delta\eta_B
}{2k})\sin(\frac{2k}{{\cal
H}_{B+}}+2k\Delta\eta_B)~,\label{largek1} \ea which implies $P_T$
is flat and oscillating rapidly with the maximal amplitude at
$k\simeq {\cal H}_{B+}$, just as showed in Fig.\ref{threep1} and
Fig.\ref{threep2}. However, if the bounce phase lasts shortly
enough, i.e. $\Delta\eta_B\sim0$, (\ref{largek1}) will be reduced
to \ba |c_{3,1}-c_{3,2}|^2\approx 1-\frac{5{\cal
H}_{B+}}{2k}\sin(\frac{2k}{{\cal H}_{B+}})+\frac{\epsilon_c{\cal
H}_{B+}}{2k}\sin(\frac{2k}{{\cal H}_{B+}})~.\label{largek2} \ea

For small $k$-modes, i.e. $k<{\cal H}_{B+}$, \ba
|c_{3,1}-c_{3,2}|^2\approx&\frac{2^{\frac{2}{1-\epsilon_c}}}{\pi}\Gamma^2(\frac{1}{2}-\frac{1}{\epsilon_c-1})(\epsilon_c-1)^{\frac{2}{\epsilon_c-1}}f(\Delta\eta_B)
(\frac{k}{{\cal H}_{B+}})^{\frac{2\epsilon_c}{\epsilon_c-1}}\sim
(\frac{k}{{\cal H}_{B+}})^{\frac{2\epsilon_c}{\epsilon_c-1}} ~,\label{smallkk} \ea which
suggests that $P_T\sim (\frac{k}{{\cal H}_{B+}})^2$ is strongly blue for $\epsilon_c\gg
1$, and $P_T\sim (\frac{k}{{\cal H}_{B+}})^3$ for $\epsilon_c\simeq 3$. The result is
consistent with that found in \cite{Cai:2015nya}. In
(\ref{smallkk}), \ba
f(\Delta\eta_B)=\Big[(1-\frac{l^2\Delta\eta_B}{2{\cal
H}_{B+}})\cosh(l\Delta\eta_B) +\frac{l}{2}(\frac{1}{{\cal
H}_{B+}}-\Delta\eta_B+\frac{l^2}{4{\cal
H}_{B+}}\Delta\eta_B^2)\sinh(l\Delta\eta_B)\Big]^2, \ea
However, if $\Delta\eta_B\sim0$,  $f(\Delta\eta_B)\sim1$ and
(\ref{smallkk}) will be reduced to \ba
|c_{3,1}-c_{3,2}|^2\approx&\frac{2^{\frac{2}{1-\epsilon_c}}}{\pi}\Gamma^2(\frac{1}{2}-\frac{1}{\epsilon_c-1})(\epsilon_c-1)^{\frac{2}{\epsilon_c-1}}(\frac{k}{{\cal H}_{B+}})^{\frac{2\epsilon_c}{\epsilon_c-1}}
~.\label{smallk2} \ea


\section{A realistic model of bounce inflation  }

\subsection{Qiu-Wang model}

How to implement the bounce before inflation has been still a
significant issue. Recently, Qiu and Wang have proposed a
realistic bounce inflation model without instabilities
\cite{Qiu:2014yn}. We briefly review it as follows.

The Lagrangian is \be {\cal
L}=\left[1-\frac{2\gamma_1}{(1+2\kappa_1\phi^2)^2}\right]X+\frac{\gamma_2
X^2 }{(1+2\kappa_2\phi^2)^2}-\frac{\gamma_3
X}{(1+2\kappa_2\phi^2)^2}\Box\phi-V(\phi), \label{L}\ee where
$X=-\partial_{\mu}\phi\partial^{\mu}\phi/2$ and \be V(\phi)=-V_0
e^{c\phi}\left[1-{\rm tanh}({{\phi}\over
\lambda_1})\right]+\Lambda^4(1-\frac{\phi^2}{v^2})^2\left[1+{\rm
tanh}({{\phi}\over \lambda_2})\right]~, \label{potential}\ee and
$M_P=1$, and the values of parameters $\gamma_1$, $\gamma_2$,
$\gamma_3$, $\kappa_1$, $\kappa_2$, $\lambda_1$, $\lambda_2$,
$V_0$, $c$, $\Lambda$ and $v$ will determine the occurrence of
bounce and inflation.

\begin{figure}[htbp]%
\begin{minipage}{\smallWidthLeft} \flushleft
\includegraphics[width=\smallWidthRight]{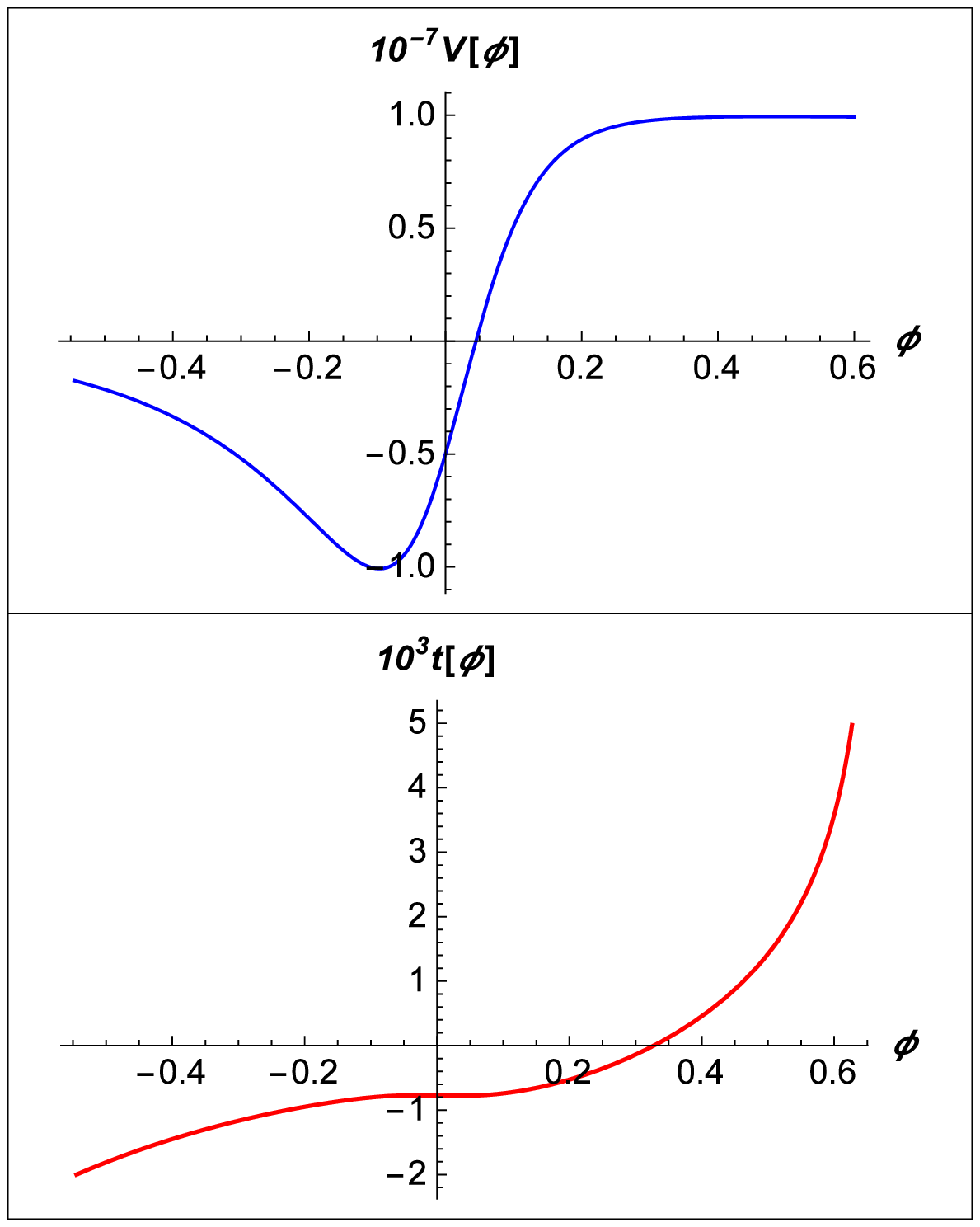}
\end{minipage}%
\begin{minipage}{\smallWidthRight} \flushleft
\includegraphics[width=\smallWidthRight]{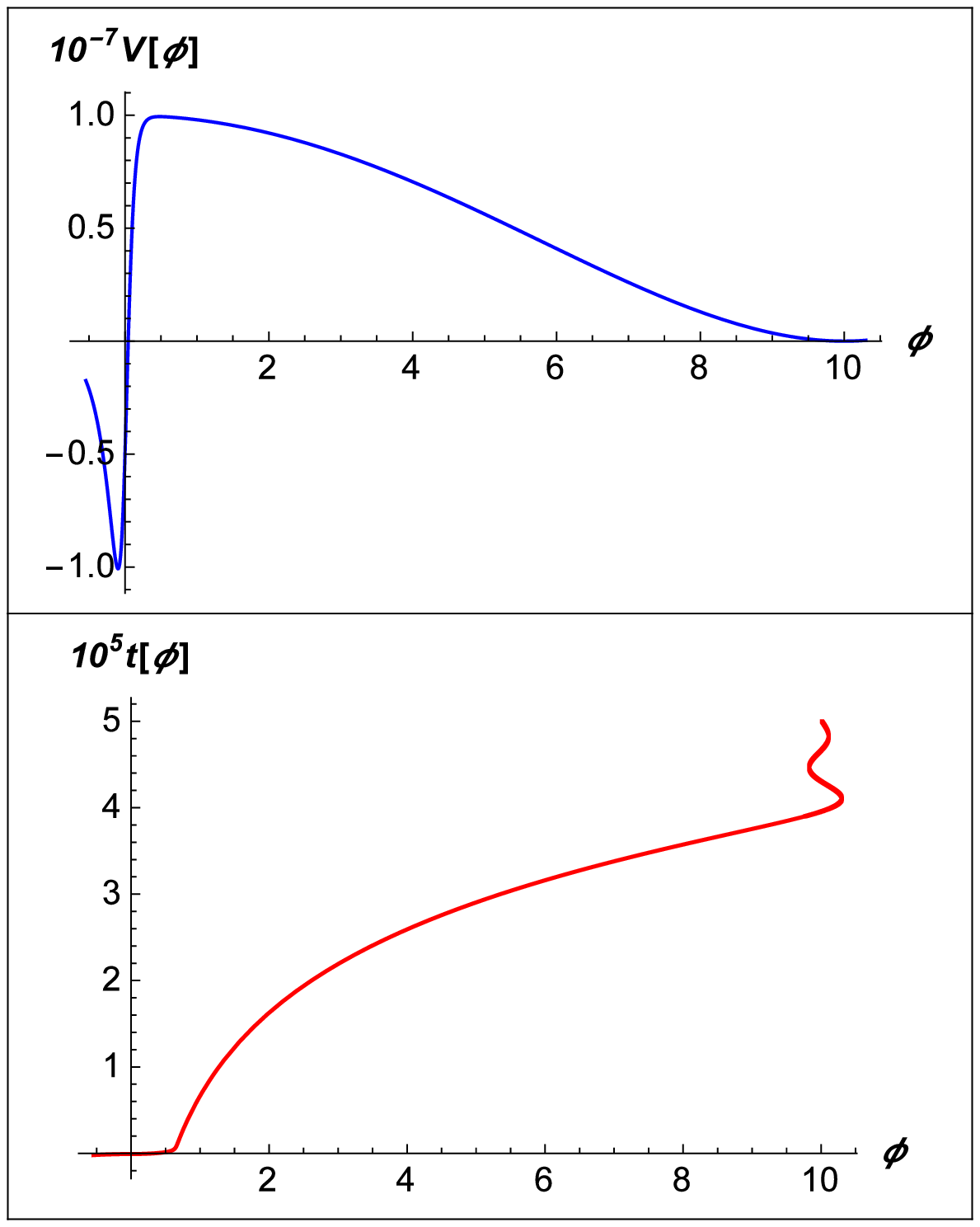}
\end{minipage}%
\caption{\label{feipol} Plots of potential $V(\phi)$ in
(\ref{potential}), and the evolution of scalar field with respect
to the physical time. }
\end{figure}

The potential is plotted in upper panels of Fig.\ref{feipol}. When
$\phi\ll -\lambda_1, 1/\sqrt{\kappa_1}, 1/\sqrt{\kappa_2}$,
(\ref{L}) is \be {\cal
L}_c=-{\partial_{\mu}\phi\partial^{\mu}\phi\over 2}+ V_0
e^{c\phi}, \ee which will bring the ekpyrotic contraction with
$\epsilon_c=c^2/2>3$. The bounce after the ekpyrotic contraction
was also studied in \cite{Osipov:2013ssa}. When $\phi\gg
-\lambda_1, 1/\sqrt{\kappa_1}, 1/\sqrt{\kappa_2}$, (\ref{L})
reduces to that of slow-roll inflation \be {\cal
L}_{inf}=-{\partial_{\mu}\phi\partial^{\mu}\phi\over
2}-\Lambda^4(1-\frac{\phi^2}{v^2})^2. \ee When $|\phi|\simeq 0$,
(\ref{L}) becomes ghost-like. However, there may be not the
instabilities, as has been confirmed in \cite{Qiu:2014yn}, see
also \cite{Libanov:2016kfc}.

\begin{figure}[htbp]
\centering
\includegraphics[scale=0.7]{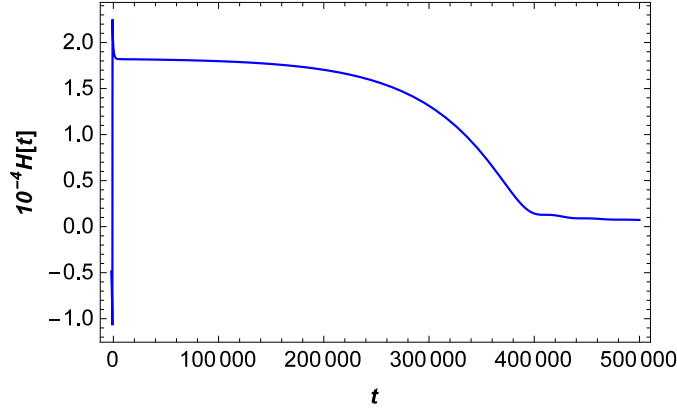}
\caption{The evolution of Hubble parameter. We set the value of
parameters as $\gamma_1=0.6$, $\gamma_2=5$, $\gamma_3=10^3$,
$\kappa_1=15$, $\kappa_2=10$, $\lambda_1=0.1$, $\lambda_2=0.1$,
$V_0=0.7$, $c=\sqrt{20}$, $\Lambda=1.5\times10^{-2}$ and $v=10$.
}\label{hubble1}
\end{figure}

We plot the evolution of background in Fig.\ref{scale1} and
Fig.\ref{hubble1}. The evolution of $\phi$ is plotted in
Fig.\ref{feipol}. Initially, we require $\phi\ll -\lambda_1,
1/\sqrt{\kappa_1}, 1/\sqrt{\kappa_2}$, and $\phi$ rolls down along
its ekpyrotic-like potential and the universe is contracting. When
$t=t_{B-}$, the ekpyrotic contraction ends, and $\phi$ climbs up
along its potential, which is essential so that the inflation can
occur subsequently, as showed in original \cite{Piao:2003zm}. When
$\phi$ arrives at $\phi\simeq 0$, the bounce will occur.
Hereafter, $\phi$ continues to climb up to the potential hill, and
then rolls slowly, the slow-roll inflation starts. The reheating
will occur around $\phi\simeq 10$, where $\phi$ oscillates and
decays.

\subsection{The spectrum of primordial GWs: Numerical result }

Here, with the background of Qiu-Wang model, which have been
plotted in Fig.\ref{scale1} and Fig.\ref{hubble1}, we will
numerically solve the perturbation equation (\ref{eom1}) and plot
the spectrum of the primordial GWs.

It is convenient for us to write the perturbation equation with
respect to the physical time \be \frac{d^2h_k}{dt^2}+3H(t)\frac{d
h_k}{dt}+\frac{k^2}{a^2(t)}h_k=0\label{phstper}, \ee and
considering $u_k=\frac{ah_k}{2}$, we have \be
\frac{d^2u_k}{dt^2}+H(t)\frac{d
u_k}{dt}+\left[\frac{k^2}{a^2(t)}u_k-H^2(t)-\frac{\ddot{a}}{a}\right]u_k=0\label{phsteru}.\ee
Initially we have
\be
u_k=\frac{1}{\sqrt{2k}}e^{-ik\int\frac{dt}{a}}, \quad
 \dot{u}_k=-i\sqrt{k\over 2 }\frac{1}{a}e^{-ik\int\frac{dt}{a}}\label{initial1} ,\ee
which suggest \be
h_k=\frac{2}{\sqrt{2k}a}e^{-ik\int\frac{dt}{a}},\quad
\dot{h}_k=-2\sqrt{2\over
k}\left(\frac{H(t)}{a}+\frac{ik}{a^2}\right)e^{-ik\int\frac{dt}{a}}\label{initial2}.\ee

\begin{figure}[htbp]
\centering
\includegraphics[scale=0.9]{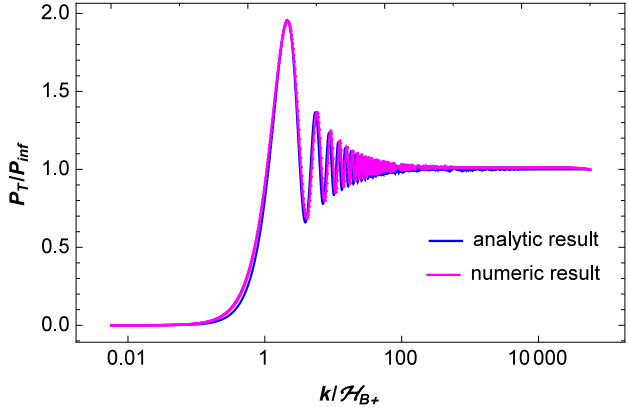}
\caption{The numeric GWs spectrum in Qiu-Wang model is compared
with our analytical result (\ref{pt2}). We set $\epsilon_c\simeq
10$, $\alpha\sim1.8\times10^{-3}{\cal H}_{B+}^2$, and
$\triangle\eta_B\sim0.2/{\cal H}_{B+}$. }\label{numeric2}
\end{figure}

We numerically plot $P_T$ in Fig.\ref{numeric2} and
Fig.\ref{numeric1}, which is completely consistent with our
analytical result (\ref{pt2}). Fig.\ref{numeric1} tells us that
the bounce phase must be short, otherwise the numerical curve will
not overlap with the analytical one. However, our (\ref{pt2}) is
actually universal, which is independent of whether the bounce is
short or not.

In addition, it should be mentioned that after the bounce, a
period with ${\dot\phi}^2$ dominated will appear, see
Fig.\ref{hubble1}. This brief period has  been often used to argue
the suppression of power spectrum, e.g.\cite{Wan:2015hya}.
However, in bounce inflation scenario, such a period is actually
only relevant with the oscillating of spectrum, while the
contraction before the bounce results in the suppression of
spectrum at large scale, as
 seen in Eqs.(\ref{largek1}) and (\ref{smallkk}).

\section{discussion }

The bounce inflation is successful in solving the initial
singularity problem of inflation and is also competitive for
explaining the power deficit in CMB at large scale. This might
provide us chance to comprehend the origin of inflation
thoroughly. There are also other designs to explain the power
deficit, such as
\cite{Contaldi:2003zv},\cite{Dudas:2012vv},\cite{BouhmadiLopez:2012by},\cite{Cicoli:2014},
but not involve the initial singularity problem. The primordial
GWs straightly encodes the evolution property of spacetime, thus
it is interesting to have a detailed study for it.

We analytically calculated the bounce inflationary GWs. The
resulting spectrum is written with the recursive Bogoliubov
coefficients including the physics of the bounce phase. The
spectrum acquires a large-scale cutoff due to the contraction,
while the oscillation around the cutoff scale is quite drastic,
which is determined by the details of bounce. We also show that
our analytic result is completely consistent with the numerical
plotting for a realistic model of bounce inflation.

In original Ref.\cite{Piao:2003zm}, the perturbation spectrum was
calculated without considering the bounce phase, which is
controversial, since conventionally it is thought that the bounce
should affect the spectrum. However, we find that if the bounce
phase lasts shortly enough, the effect of the bounce on the
primordial GWs may be negligible, and the calculation without
considering the bounce phase is robust.

We should point out that what we applied is the perturbation
equation without modifying gravity, which may be implemented only
when the null energy condition is broke. However, the physics of
bounce phase is unknown, it is obvious that the high-curvature
corrections of gravity may also result in the occurrence of
bounce, e.g. the Gauss-Bonnet correction
\cite{Bamba:2014mya},\cite{Haro:2015oqa}, and non-local gravity
\cite{Biswas:2011ar}, which will inevitably modify the
perturbation equation. We find that the corresponding corrections
will aggravate the oscillating behavior around the cutoff scale
${\cal H}_{B+}$, however, the spectrum in the regime of $k\ll
{\cal H}_{B-}$ and $k\gg {\cal H}_{B+}$ is hardly affected. We
will come back to this issue in upcoming work.

In addition, as has been mentioned, the bounce inflation may also
explain a large dipole power asymmetry in CMB at low-l. However,
the asymmetry might also appear in CMB B-mode polarization
\cite{Abolhasani:2013vaa},\cite{Namjoo:2014pqa}. Moreover, during
the bounce, there might be a large parity violation
\cite{Wang:2014abh}. It is interesting to have a reestimate for
the relevant issues.

Series of experiments aiming at detecting GWs have been
implemented or will be implemented. Our work suggests that
searching primordial GWs at large scale might tell us if the
bounce has ever happened before inflation.

\textbf{Acknowledgments}

This work is supported by NSFC, No. 11222546, 11575188, and the
Strategic Priority Research Program of Chinese Academy of
Sciences, No. XDA04000000.

\appendix

\section{The effects of $u_k$ continuum and $h_k$ continuum on spectrum  }

When we do calculations, it is convenience to write $h_k$ as
${u_k}/{2a(\eta)}$, since the evolution of $u_k$ satisfies a
Bessel equation. However, the real GWs mode is $h_k$, not ${u_k}$.
Thus which of $u_k$ or $h_k$ is continuous on the matching surface
might be a question needed to be clarified. Our result (\ref{pt2})
is based on the continuum of $h_k$, which we will argue as
follows.

We define the phase `$i$' and `$i+1$' as the conjoint phases, both
are matched at $\eta_0$. The perturbations in different phase
satisfy \ba \label{hiui}\left(
\begin{array}{ccc} h_{i}\\h_{i}^{'}
\end{array}\right)&=&\left(\begin{array}{ccc} \frac{1}{a_i}&0\\-\frac{a_i^{'}}{a_i^2}&\frac{1}{a_i}\end{array}\right) \left(\begin{array}{ccc}u_{i}\\u_{i}^{'}
\end{array}\right)~,
\ea and \ba \label{hipui}\left(
\begin{array}{ccc} h_{i+1}\\h_{i+1}^{'}
\end{array}\right)&=&\left(\begin{array}{ccc} \frac{1}{a_{i+1}}&0\\-\frac{a_{i+1}^{'}}{a_{i+1}^2}&\frac{1}{a_{i+1}}\end{array}\right) \left(\begin{array}{ccc}u_{i+1}\\u_{i+1}^{'}
\end{array}\right)~,
\ea where the prefactor 2 is neglected. The the continuum of $h_k$
and ${\dot h}_k$ at the matching surface $\eta_0$ suggest \ba
\label{hipuip}\left(
\begin{array}{ccc} h_{i}\\h_{i}^{'}
\end{array}\right)&=&\left(
\begin{array}{ccc} h_{i+1}\\h_{i+1}^{'}
\end{array}\right)~.
\ea Thus we have, \ba \left(\begin{array}{ccc}u_{i+1}\\u_{i+1}^{'}
\end{array}\right)&=&\left(\begin{array}{ccc} \frac{1}{a_{i+1}}&0\\-\frac{a_{i+1}^{'}}{a_{i+1}^2}&\frac{1}{a_{i+1}}\end{array}\right)^{-1}
\left(\begin{array}{ccc} \frac{1}{a_i}&0\\-\frac{a_i^{'}}{a_i^2}&\frac{1}{a_i}\end{array}\right) \left(\begin{array}{ccc}u_{i}\\u_{i}^{'}
\end{array}\right)~.
\ea Finally, \be u_{i+1}=\frac{a_{i+1}}{a_{i}}u_{i}~,\quad
u_{i+1}^{'}=\left(\frac{a_{i+1}^{'}}{a_{i}}-\frac{a_{i+1}^{}a_{i}^{'}}{a_{i}^2}+\frac{a_{i+1}}{a_{i}}\right)u_{i}^{'}~.
\ee Generally, $a_{i+1}=a_{i}$ at the matching surface. Thus if
$a_{i+1}^{'}=a_{i}^{'}$, the continuum of $u$ is equal to that of
$h$.

\begin{figure}[htbp]
\centering
\includegraphics[scale=0.7]{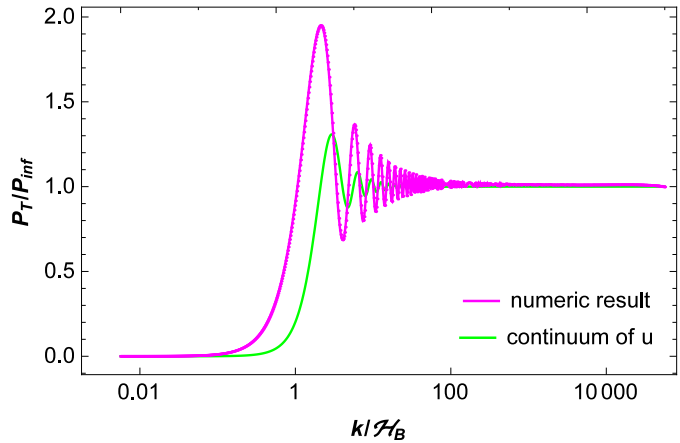}
\caption{The numeric result in Sec.IV is compared with the
analytical result, which is calculated by considering the
continuity of $u_k$, with $\epsilon_c\simeq 10$. }\label{ucon}
\end{figure}

In Sec.III.D, $a_{i+1}^{'}=a_{i}^{'}$ is assured. We plot $P_T$
with the continuum of $u$ and $h$, respectively, in
Fig.\ref{hcon}, which are completely identical. However, in
Appendix B, the bounce phase is neglected, and we have $a^{'}(\eta_{B+})\simeq
-a^{'}(\eta_{B-})$, which indicates that $a^{'}$ is not continuous. We analytically calculate
$P_T$ with the continuum of $u$, and plot it in Fig.\ref{ucon}. It
is found that the analytical result obtained can't match with the
numeric curve accurately.

\begin{figure}[htbp]
\centering
\includegraphics[scale=0.7]{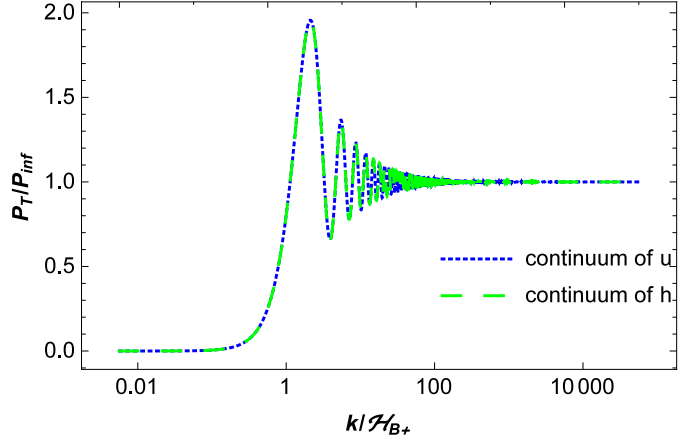}
\caption{The results are obtained by considering the continuity of
$h_k$(green curve) and  $u_k$(blue curve), with $\epsilon_c\simeq
10$, $\alpha\sim1.8\times10^{-3}{\cal H}_{B+}^2$, and
$\triangle\eta_B\sim0.2/{\cal H}_{B+}$. }\label{hcon}
\end{figure}

\section{Is the effect of bounce phase negligible ?}


In original idea of bounce inflation \cite{Piao:2003zm}, the
primordial GWs spectrum was calculated by straightly jointing the
perturbation solution in the contracting phase to that in the
expanding phase. It is interesting to estimate if the effect of
bounce phase negligible.

We do not consider the bounce phase, which implies $\Delta\eta=0$
and ${\cal H}_{B+}={\cal H}_{B-}={\cal H}_B$. The spectrum of
primordial GWs is still Eq.(\ref{pt2}). However, $c_{3,1}$ and
$c_{3,2}$ should be determined by straightly matching the
solutions (\ref{contract}) and (\ref{inflation}). The perturbation
$\gamma_{ij}$ and its time derivative should be continuous through
the match surface. Thus we have \ba \label{bb}\left(
\begin{array}{ccc} c_{3,1}\\c_{3,2}
\end{array}\right)&=&{\cal
N}^{(3,1)}\times\left(\begin{array}{ccc}c_{1, 1}\\c_{1,
2}\end{array}\right)\label{rec} \ea with the matric ${\cal
N}^{(3,1)}$ \ba {\cal N}_{11}^{(3,1)}&=&\frac{i\pi k \sqrt{x_1
x_2} }{8}\bigg\{[H^{(1)}_{\nu_{1}+1}(k x_1)-H^{(1)}_{\nu_{1}-1}(k
x_1)]H^{(2)}_{\nu_{2}}(k x_2) \nonumber\\
&+&[H^{(2)}_{\nu_{2}-1}(k x_2)-H^{(2)}_{\nu_{2}+1}(k
x_1)]H^{(1)}_{\nu_{1}}(k x_1)
+\frac{2}{k}(\frac{\nu_{2}}{x_2}-\frac{\nu_{1}}{x_1})H^{(1)}_{\nu_{1}}(k
x_1)H^{(2)}_{\nu_{2}}(k x_2) \bigg\},
\nonumber\\
{\cal N}_{12}^{(3,1)}&=&\frac{i\pi k \sqrt{x_1 x_2}
}{8}\bigg\{[H^{(2)}_{\nu_{1}+1}(k x_1)-H^{(2)}_{\nu_{1}-1}(k
x_1)]H^{(2)}_{\nu_{2}}(k x_2)\nonumber\\ &+&[H^{(2)}_{\nu_{2}-1}(k
x_2)-H^{(2)}_{\nu_{2}+1}(k x_1)]H^{(2)}_{\nu_{1}}(k x_1)
+\frac{2}{k}(\frac{\nu_{2}}{x_2}-\frac{\nu_{1}}{x_1})H^{(2)}_{\nu_{1}}(k
x_1)H^{(2)}_{\nu_{2}}(k x_2) \bigg\},
\nonumber\\
{\cal N}_{21}^{(3,1)}&=&\frac{i\pi k \sqrt{x_1 x_2}
}{8}\bigg\{[H^{(1)}_{\nu_{1}-1}(k x_1)-H^{(1)}_{\nu_{1}+1}(k
x_1)]H^{(1)}_{\nu_{2}}(k x_2)\nonumber\\
&+&[-H^{(1)}_{\nu_{2}-1}(k x_2)
+H^{(1)}_{\nu_{2}+1}(k x_2)]H^{(1)}_{\nu_{1}}(k
x_1)+\frac{2}{k}(\frac{\nu_{1}}{x_1}-\frac{\nu_{2}}{x_2})H^{(1)}_{\nu_{1}}(k
x_1)H^{(1)}_{\nu_{2}}(k x_2) \bigg\},
\nonumber\\
{\cal N}_{22}^{(3,1)}&=&\frac{i\pi k \sqrt{x_1 x_2}
}{8}\bigg\{[H^{(2)}_{\nu_{1}-1}(k x_1)-H^{(2)}_{\nu_{1}+1}(k
x_1)]H^{(1)}_{\nu_{2}}(k x_2)\nonumber\\
&+&[-H^{(1)}_{\nu_{2}-1}(k x_2)
+H^{(1)}_{\nu_{2}+1}(k x_2)]H^{(2)}_{\nu_{1}}(k x_1)
+\frac{2}{k}(\frac{\nu_{1}}{x_1}-\frac{\nu_{2}}{x_2})H^{(2)}_{\nu_{1}}(k
x_1)H^{(1)}_{\nu_{2}}(k x_2) \bigg\}. \nonumber \ea For large
$k$-modes, i.e. $k>{\cal H}_{B}$, \ba |c_{3,1}-c_{3,2}|^2
\approx&1-\frac{5{\cal H}_{B}}{2k}\sin(\frac{2k}{{\cal
H}_{B}})+\frac{\epsilon_c{\cal H}_{B}}{2k}\sin(\frac{2k}{{\cal
H}_{B}})~,\label{largek3} \ea which is correspond to
(\ref{largek1}). While for small $k$-modes, i.e. $k<{\cal H}_{B}$,
the result is same with (\ref{smallkk}).

\begin{figure}[htbp]
\centering
\includegraphics[scale=1]{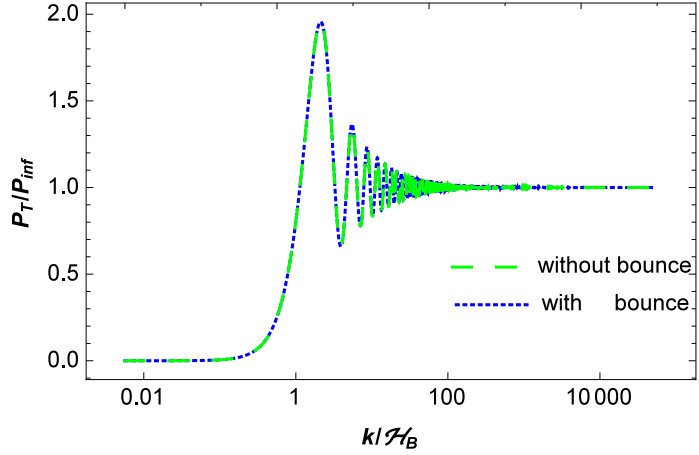}
\caption{ The spectrum (\ref{pt2}) is compared with that without
considering the bounce phase. We set $\epsilon_c\simeq 10$,
$\alpha\sim1.8\times10^{-3}{\cal H}_{B+}^2$, and
$\triangle\eta_B\sim0.2/{\cal H}_{B+}$. }\label{threep3}
\end{figure}

\begin{figure}[htbp]
\centering
\includegraphics[scale=1]{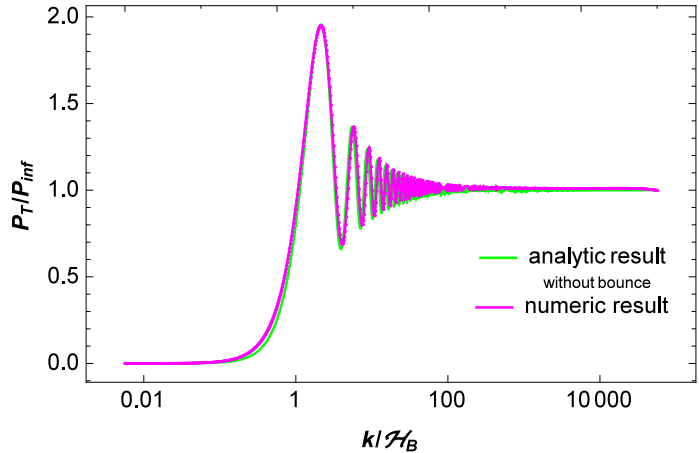}
\caption{The numeric GWs spectrum in Sec.IV is compared with our
analytical result (\ref{rec}) without considering the bounce
phase. $\epsilon_c\simeq 10$. }\label{numeric1}
\end{figure}

In Fig.\ref{threep3}, we plot $P_T$ with (\ref{rec1}) and
(\ref{rec}), respectively. Here, $\epsilon_c\simeq 10$. We see
that if $\triangle\eta_B\sim 0.2/{\cal
H}_{B+}$ is set to parameterise the
physics of bounce, both curves completely overlap. This indicates
that if the bounce lasts shortly enough, the effect of the bounce
on the primordial GWs may be negligible. It is noticed in Sec.III
that in realistic model of bounce inflation, the period of bounce
is actually short enough, thus the result without considering the
bounce phase is robust, see Fig.\ref{numeric1}.

\end{document}